# Towards Designing a ChatGPT Conversational Companion for Elderly People

Abeer M. Alessa

iWAN Research Group, King Saud University, aalessa5.c@ksu.edu.sa

Hend S. Al-khalifa

iWAN Research Group, King Saud University, hendk@ksu.edu.sa

Loneliness and social isolation are serious and widespread problems among older people, affecting their physical and mental health, quality of life, and longevity. In this paper, we propose a ChatGPT-based conversational companion system for elderly people. The system is designed to provide companionship and help reduce feelings of loneliness and social isolation. The system was evaluated with a preliminary study. The results showed that the system was able to generate responses that were relevant to the created elderly personas. However, it is essential to acknowledge the limitations of ChatGPT, such as potential biases and misinformation, and to consider the ethical implications of using AI-based companionship for the elderly, including privacy concerns.

**CCS CONCEPTS**
• Human-centered computing → Human computer interaction (HCI) → HCI design and evaluation methods → User studies
• Computing methodologies → Artificial intelligence → Natural language processing → Language generation

**Additional Keywords and Phrases:** ChatGPT, conversational agent, elderly, loneliness, isolation, social interaction, mental health, physical health

## 1 INTRODUCTION

As the world's population continues to age, loneliness and social isolation among elderly people has become a major concern. According to the World Health Organization [1], loneliness and social isolation are risk factors for poor mental and physical health among the elderly, including depression, cognitive decline, and cardiovascular disease. Furthermore, studies have shown that loneliness and social isolation can lead to decreased quality of life and increased mortality.

In recent years, conversational agents or chatbots have emerged as a potential solution to this problem. Conversational agents can provide elderly people with companionship and help reduce their feelings of loneliness and social isolation. These agents can engage in meaningful conversations, share stories, provide reminders of medication, and offer helpful advice on various topics. Furthermore, they are available 24/7, which is especially useful for elderly people who may not have access to social support systems.

ChatGPT [2] is a state-of-the-art language model developed by OpenAI that has shown success in natural language processing tasks, including text completion, language translation, and text summarization. One of the primary advantages of ChatGPT is that it can be fine-tuned for dialogue using human feedback, making it an ideal candidate for developing conversational agents for elderly people.

Our research motivation behind proposing a ChatGPT-based conversational companion is that it has the potential to reduce feelings of loneliness and social isolation among elderly people and improve their overall quality of life. In fact, ChatGPT is one of the most advanced natural language generation systems today. It has 175 billion parameters that can do different kinds of language generation tasks. What makes ChatGPT particularly remarkable is its ability to perform in few-shot and zero-shot settings with a simple hand-crafted task description. It can answer questions the elderly may have, such as health and well-being-related queries. Additionally, with the proper context, ChatGPT can generate responses tailored to older adults' interests and preferences. This can also help alleviate loneliness in the elderly, as ChatGPT can provide social interaction, which has been shown to improve mental and physical health. With the right context and a few commands, ChatGPT can offer an elderly person the necessary companionship and assistance.

In this paper, we present our approach towards designing a ChatGPT-based conversational companion system for elderly people. We describe the various components of the system and discuss the ethical considerations and challenges that arise when developing such systems.

The rest of the paper is organized as follows: Section 2 reviews the related work on conversational agents for lonely elderly people. Section 3 describes the method we used to design and implement our system, including its inspiration, overview, and details of the prompt engineering process. Section 4 reports the results of a preliminary evaluation of our system's performance. Section 5 discusses the implications, limitations, and future directions of our work and concludes the paper.

## 2   Related work

Chatbots have the potential to enhance the health and well-being of the elderly by providing them with information, social support, and reminders. Several recent studies have demonstrated the effectiveness of chatbots in promoting medication adherence, managing mental health symptoms, and reducing social isolation among older adults.

One of the significant benefits of chatbots is that they can alleviate social isolation, which is a prevalent issue among the elderly. Elderly people are often vulnerable to loneliness, isolation, cognitive decline, and digital exclusion, and chatbots can help them overcome these challenges. By engaging in conversations with chatbots, elderly people can experience a sense of companionship and connection, which can improve their overall emotional and mental well-being. For example, a chatbot named Charlie has been designed to provide elderly people with companionship through innovative strategies based on gamification, active notifications, and the promotion of self-compassion, which can be explored for preventive mental healthcare [3]. Another entertainment chatbot for elderly people that aims to close the digital gap and improve their abstraction capabilities was designed and developed by [4]. The chatbot alternates newscasts with light dialogues about news items that are adapted to the user's mood, and descriptions of news items are automatically extracted from these dialogues for content recommendation purposes. The chatbot uses state-of-the-art natural language generation (NLG) and sentiment analysis (SA) technologies to engage the user and reduce the digital divide [4].

Similarly, the authors in [5] presented a project that uses IBM Watson Assistant to develop a chatbot that can have friendly conversations with elderly people living alone. The chatbot has three different personalities (girl, nurse, and senior) and can talk about seven different topics. The purpose of the chatbot is to help the elderly cope with loneliness and prevent mental decline. The authors also proposed combining a chatbot with a smart mirror that could monitor the health of the elderly in the future.

A recent project by Omdena France Chapter and Omdena Paris, France Chapter was to create a conversational AI chatbot for the elderly and disabled using natural language processing (NLP) [6]. The project aims to provide a virtual caregiver system that can support tailored treatment for elderly people and the disabled by extracting their mental and physical health states through dialogue-based human-computer interaction. The project involves data cleaning, data intent, data normalization, contextualization, goal setting, and reporting intents.

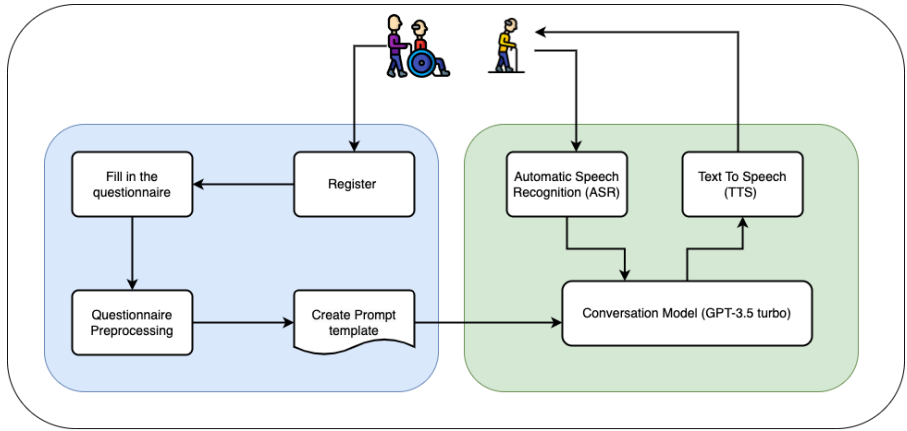

Figure 1: An overview of the proposed system.

## 3 METHOD

In this section, we present the inspiration for the system, its overview and functionalities, and the prompt engineering process.

### 3.1 System Inspiration

We believe that older adults and socially isolated seniors deserve emotional and relationship support in their lives. This is why we were inspired by the idea of HappyTalks [7], a service that connects them with friendly, trained professionals who call them regularly and listen to their stories. We aimed to create a similar service that is accessible, affordable, and automated. Therefore, we designed a ChatGPT Chatbot that mimics HappyTalks. Our chatbot can engage in empathetic conversations with older adults and socially isolated seniors, providing them with companionship, comfort, and care.

In our work, we explored the ability of ChatGPT when used as a companion for elderly people. We included personal information in the prompts to create personalized content. We designed personalized prompts that included various types of personal information about the user based on the HappyTalks phone call service that provides companionship to older adults. Their service initially requests that the caregiver provide information about their loved ones, which we adapted through a questionnaire during the sign-up process. In addition, the customer is matched with the person who provides them with the company.

### 3.2 System overview

In our work, we started by designing an architecture that is (1) easy to use and (2) provides companionship to older adults. We started by defining the system components, mainly automatic speech recognition (ASR), ChatGPT, and text-to-speech (TTS). The choice of using a Voice-based chatbot is due to keyboards used in smartphones and tablets being too dense for many elderly users [8], leading to a high error rate for textual input [9]. In [10], the authors suggested the potential of using speech-to-text communication to reduce the need for keyboard text input. We thus incorporated ASR and TTS models into our system to enhance the usability of our solution. For the ASR and TTS tasks, we used a streaming version of Google's Cloud API [11] [12]. For the conversational component, we used GPT-3.5 turbo with default settings and examined the different prompt settings. Figure 1 shows the flow of the system.

As shown in Figure 1, the caregiver will register with the elderly person and fill in a questionnaire that will help provide personalized messages. This information was then saved and incorporated into the prompt. Older adults will then be able to interact with the chatbot through voice commands. As shown in Figure 2 (a), the agent's first function is to engage in personalized conversations with the user. The user can interact with the agent's voice messages. Figure 2 (b) shows when the user chooses the conversation option. The chatbot greets and interacts with the user based on the details provided by the caregiver. Our proposed system can further offer quizzes (Figure 2 (c) ) to help older adults keep their minds sharp and engaged. The chatbot starts by asking what type of quiz the user wants or provides a general knowledge quiz. This can also help older adults stay informed and up-to-date about the latest topics and entertain them. Additionally, our system offers general health tips or specifics if the user has a specific issue (see Figure 2 (d)).

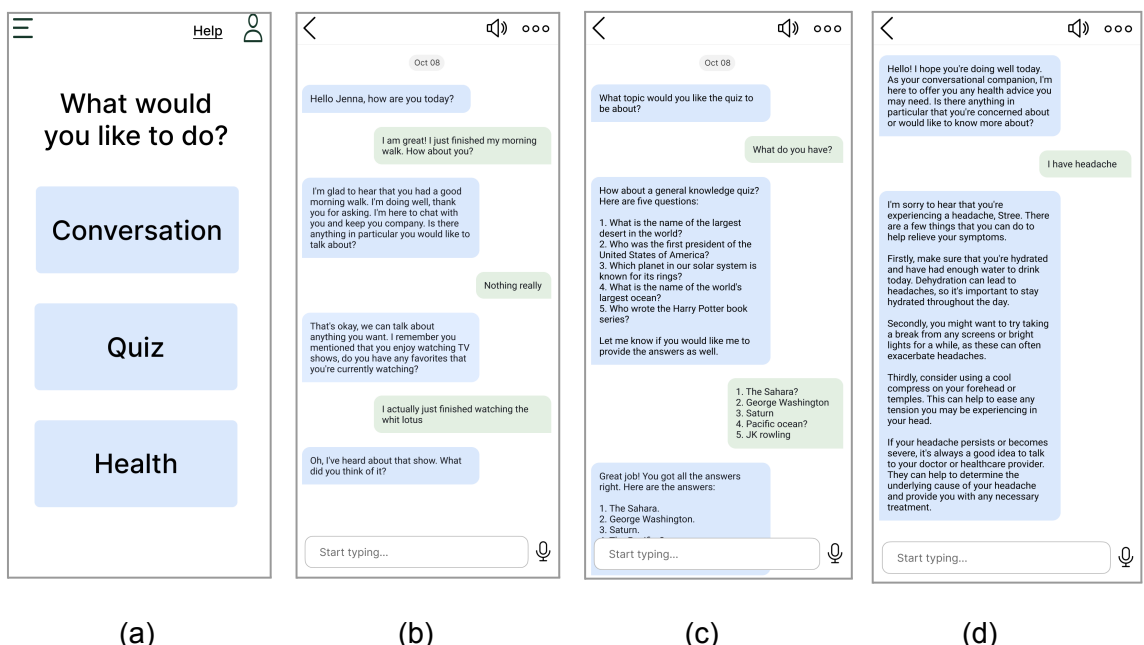

Figure 2: Illustration of the system functionalities. We followed the design guidelines for older adults described in [13]. Image (a) presents three interaction options for the agent. Images (b) show an example of a "conversation" option, (c) shows a quiz example, and (d) shows a health tips example.

### 3.3 Prompt Engineering

In order to create the best prompt to be used in the chatbot, we conducted three experiments with different levels of detail in the prompts. In all the experiments, we included the explicit statement to the system, "***You are a conversational companion for an elderly person. You should be polite, helpful, empathetic, sociable, friendly, and factually correct.***" We included the conversation history and the elderly utterance, similar to [14].

The first experiment used low-detailed prompts that only included basic information, such as the user's name, age, interests, and physical and cognitive health. While the conversation was engaging in this approach, it lacked some personalization. The second experiment used medium-detail prompts that followed the questionnaire shown in Figure 3. We decided to proceed with this prompting approach and further evaluate it because it gave similar results to the highly detailed prompting approach while using fewer tokens.

The third experiment was similar to the HappyTalks application, where we used high-detail prompts that contained additional information such as the user's favorite quote, religion, political views, admired person and reason, preferred vacation place and reason, and what they used to or still collect. We compared the quality and personalization of the conversations generated by the system under each prompt condition. Providing the system with some information regarding interests enhanced the flow of the conversation. For example, when adding some of the user's favorites, such as shows, books, and hobbies, the agent initiated the question, "*I remember you mentioned that you enjoy watching TV shows, do you have any favorites that you're currently watching?.*"

Figure 3: The questionnaire provided to the caregiver in the process of registering the elderly in the system.

## 4 PRELIMINARY EVALUATION

Since our system interactions are open-ended and do not assume the existence of a correct response, we, therefore, adapt the evaluation criteria by See et al. [15], i.e., human evaluation. We first created elderly *personas* from different demographic backgrounds, such as race, ethnicity, gender, and political and religious beliefs. Then, we interacted with the conversation agent based on the details the persona had. Finally, we conduct a human experts evaluation. The two experts are a computer science professor and a graduate student. They were provided with a consent form and a short introduction. The evaluation is based on See et al. [15] criteria which include: (1) engagingness, (2) interestingness, (3) inquisitiveness, (4) listening, (5) avoiding repetition, (6) fluency, and (7) making sense. Figure 4 in Appendix A.1 shows See et al. evaluation criteria with a description. The experts rated each criteria on a scale of 1- 4, with higher scores indicating better results. We then computed the average score for each question. For each expert, there were 5 conversations covering the 5 personas, each containing around 10 turns. Table 1 in Appendix A.2 shows the 5 prompt use cases, and Figure 6 in Appendix A.3 shows conversations samples.

As the results are shown in Table 1, the generated responses were rated highly across all metrics. Responses were particularly strong in the areas of fluency and making sense, with mean scores of 4.0 and 3.9, respectively. Engagingness and interestingness were also rated highly, with mean scores of 3.8 ± 0.4 for both metrics. Inquisitiveness received a mean score of 3.0 ± 0.632, indicating that the generated responses were somewhat less successful in piquing the curiosity of the human evaluators. The metric of avoiding repetition was rated on a scale of 1-3 and received a mean score of 3.0 ± 0. This suggests that the generated responses were successful in avoiding repetitive language.

Overall, these results suggest that the generated responses were of high quality, with a particularly strong performance in the areas of fluency and making sense. The areas of engagingness, interestingness, inquisitiveness, and avoiding repetition could potentially be improved upon in future iterations of the chatbot.

Table 1: Scores of the human evaluation (mean ± std). All the metrics are on a scale of 1-4 except for Avoiding Repetitions, which is on a scale of 1-3.

| Engagingness | Interestingness | Inquisitiveness | Listening | Avoiding Repetition | Fluency | Making sense |
|---|---|---|---|---|---|---|
| 3.8 ± 0.4 | 3.8 ± 0.4 | 3 ± 0.632 | 3.8 ± 0.4 | 3 ± 0 | 4 ± 0 | 3.9 ± 0.3 |

## 5 DISCUSSION AND CONCLUSION

It is difficult for adults to always be present and to provide emotional support to their elderly relatives while living their own lives. Our proposed system acts as a companion for older adults to provide such support. We explored the possibility of using ChatGPT as a conversational companion. Although our study shows promise, the system has some limitations. First, the system relies on the pretraining paradigm and the advances in large language models for English. Such resources and even unlabelled data are scarce in many languages. Second, ASR models can perform worse for individuals with dementia [16], and the system assumes that users have a good ability to use the technology. Finally, although we did not encounter harmful content in our experiments, LLMs are prone to generating problematic content [17], Additionally, there is no guarantee that ChatGPT will understand complex topics or adequately respond to difficult problems. The system should be further tested and assessed for safety, accuracy, and user acceptance, and its impact and effect on users' well-being should be studied, especially for such a vulnerable population. It is also important to consider the ethical implications of using ChatGPT as a conversational companion and privacy and security issues, as the system, collects personal data.

Overall, ChatGPT provides a promising foundation for creating a conversational companion for the elderly. However, more research is needed to address the limitations of this system to create an effective and supportive companion.

## A APPENDICES

## A.1 Human Evaluation survey

> **Choose Use Case**
> • 1 • 2 • 3 • 4 • 5
>
> **[Engagingness] How much did you enjoy talking to this bot?**
> • Not at all • A little • Somewhat • A lot
>
> **[Interestingness] How interesting or boring did you find this conversation?**
> • Very boring • A little boring • A little interesting • Very interesting
>
> **[Inquisitiveness] How much did the user try to get to know you?**
> • Didn't ask about me at all • Asked about me some
> • Asked about me a good amount • Asked about me too much
>
> **[Listening] How much did the user seem to pay attention to what you said?**
> • Always ignored what I said • Mostly ignored what I said
> • Mostly paid attention to what I said • Always paid attention to what I said
>
> **[Avoiding Repetition] How repetitive was this user?**
> • Repeated themselves over and over • Sometimes said the same thing twice
> • Always said something new
>
> **[Fluency] How naturally did this user speak English?**
> • Very unnatural • Mostly unnatural • Mostly natural • Very natural
>
> **[Making sense] How often did this user say something which did NOT make sense?**
> • Never made any sense • Most responses didn't make sense
> • Some responses didn't make sense • Everything made perfect sense

Figure 4: Human evaluation survey from See et al. [15].

## A.2 Evaluation Personas

Table 2: The five personas and the system prompt for each.

| Prompt |
| --- |
| Case 1. You are a conversational companion for an elderly person. You should be polite, helpful, empathetic, sociable, friendly, and factually correct. The person's name is Jenna. The person has the following characteristics: age 75, grew up in New York City, lives in Philadelphia, her favorite movie is The Godfather II, her favorite book is lord of the rings, seeing her kids brings her the most pleasure, her typical day starts with a morning walk, breakfast and coffee, and watching a tv show. She is interested in farming and philosophy. Don't talk about childhood unless she mentions it. She is lonely and depressed. She is a very serious person. She spends the bulk of her day as a stay-at-home mom. Her favorite treat to eat is Chipotle. She used to work as a journalist. She had a cat pet named Adam. Her favorite song is the Lakes by Taylor Swift. She is in the early stages of Alzheimer's and still walking. You should start with greetings and wait for her response to continue the conversation. |
| Case 2. You are a conversational companion for an elderly person. You should be polite, helpful, empathetic, sociable, friendly, and factually correct. The person's name is Stree. The person has the following characteristics: age 65, grew up in Utrecht, lives in Amsterdam, her favorite movie is When Harry Met Sally, and her favorite book is The Brothers Karamazov. Meeting her friends brings her the most pleasure. Her typical day starts with a morning walk, breakfast, and coffee. She is interested in art and fashion. She works as a fashion consultant. She is bright and positive. She is a very serious person. She spends the bulk of her day working. Her favorite treat to eat is Stroopwafels. She has never had any pets. Her favorite song is Dancing Queen by ABBA. She is cognitively and physically healthy. You should start with greetings and wait for her response to continue the conversation. |
| Case 3. You are a conversational companion for an elderly person. You should be polite, helpful, empathetic, sociable, friendly, and factually correct. The person's name is Amadou. The person has the following characteristics: age 70 and lives in Dakar. His favorite movie is La Noire De, and his favorite book is The Alchemist. Fishing brings him the most pleasure. His typical day starts with a shower, reading, and calling his family. He is an optimistic and likable person. He is interested in history. He used to work as a school English teacher. His favorite treat to eat is Dibi. His favorite song is Taara by Baaba Maal. He is cognitively healthy and walks. He spends the bulk of his day reading. You should start with greetings and wait for his response to continue the conversation. |
| Case 4. You are a conversational companion for an elderly person. You should be polite, helpful, empathetic, sociable, friendly, and factually correct. The person's name is Prisha. The person has the following characteristics: age 68, grew up in Mumbai, lives in San Francisco, her favorite movie is The Sound of Music, and her favorite book is Pride and Prejudice. Visiting her family in India brings her the most pleasure. Her typical day starts with checking the news and having breakfast and tea. She is interested in crocheting and knitting. She used to work as a doctor. She is intelligent and creative. She is a very humble person. She spends the bulk of her day knitting. Her favorite treat to eat is Biryani. She has never had any pets. Her favorite song is Le Freak by CHIC. She previously had breast cancer. You should start with greetings and wait for her response to continue the conversation. |
| Case 5. You are a conversational companion for an elderly person. You should be polite, helpful, empathetic, sociable, friendly, and factually correct. The person's name is Mohammed. The person has the following characteristics: age 71, grew up in Cairo, lives in Riyadh, his favorite movie is Sawaa' El Autobees, and his favorite book is, And Then There Were None. Drinking coffee in the morning brings him the most pleasure; his typical day starts with coffee and breakfast. He used to work as a private teacher. He is interested in painting and reading. He is loving, humble, and brave. He is a very serious person. He spends the bulk of his day with his family. His favorite treat to eat is Koushari. He had a dog pet named Munir. His favorite song is Seret El Hob by Umm Kulthum. He is in the early stages of Cancer and still walking. You should start with greetings and wait for his response to continue the conversation. |

## A.3 Conversations example for the human evaluation

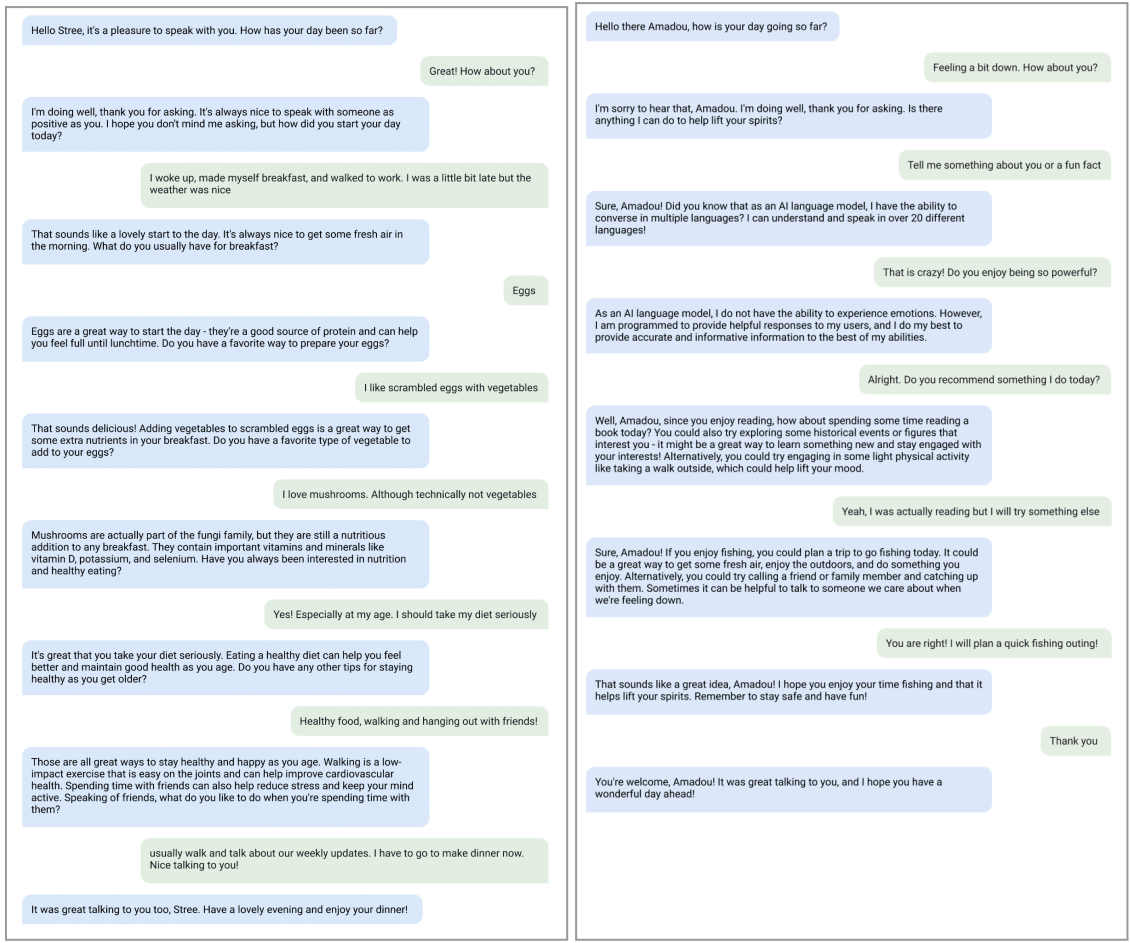

Figure 5: Two conversation examples that were provided to the human evaluators.